# Room Temperature Formation of Carbon Onions via Ultrasonic Agitation of MoS$_2$ in Isopropanol

Andrew J. Stollenwerk. Eric Clausen, Matthew Cook, Keith Doore, Ryan Holzapfel, Jacob Weber, Rui He, and Timothy E. Kidd [*]

*University of Northern Iowa, Department of Physics, 215 Begeman Hall, Cedar Falls, Iowa 50614-0150*

∗ Corresponding author.

E-mail address: tim.kidd@uni.edu

Phone: 319-2732380

Mailing Address: 304 Begeman Hall, University of Northern Iowa, Cedar Falls, IA, US, 50614-0150

**Abstract**

Ultrasonic agitation is a proven method for breaking down layered materials such as $MoS_2$ into single or few layer nanoparticles. In this experiment, $MoS_2$ powder is sonicated in isopropanol for an extended period of time in an attempt to create particles of the smallest possible size. As expected, the process yielded a significant quantity of nanoscale $MoS_2$ in the form of finite layer sheets with lateral dimensions as small as a few tens of nanometers. Although no evidence was found to indicate a larger the longer sonication times resulted in a significant increase in yield of single layer $MoS_2$, the increased sonication did result in the formation of several types of carbon allotropes in addition to the sheets of $MoS_2$. These carbon structures appear to originate from the breakdown of the isopropanol and consist of finite layer graphite platelets as well as a large number of multi-walled fullerenes, also known as carbon onions. Both the finite layer graphite and $MoS_2$ nanoplatelets were both found to be heavily decorated with carbon onions. However, isolated clusters of carbon onions could also be found. Our results show that liquid exfoliation of $MoS_2$ is not only useful for forming finite layer $MoS_2$, but also creating carbon onions at room temperature as well.

## 1. Introduction

Research into layered materials has reached a new apex with the discovery of emergent properties as these systems are formulated into single or finite layer constructs. While the most famous example is graphene [1], the transition metal dichalcogenide $MoS_2$ is also seeing a high level of interest. Sharing the high mobility characteristics of graphene, single layer $MoS_2$ has a direct band gap making it suitable for transistor or electro-optical applications [2]. $MoS_2$ is stable in ambient conditions, sufficiently inexpensive to be used as a common lubricant and readily available naturally or via chemical synthesis [3]. As such, it is one of the most promising transition metal dichalcogenides for use in commercial or industrial applications. However, there is still much work to be done to reliably and economically break down or synthesize $MoS_2$ into finite or single layer form. While isolated single layers can be achieved using the scotch tape method or simply by rubbing a crystal up against a substrate [4], such methods are intrinsically unsuitable for mass production.

One method which does show promise for mass production, liquid exfoliation, is attractive in that it can be used to break down a relatively large amount of $MoS_2$ into finite layer form [5]. This technique breaks down $MoS_2$ in an organic solvent via cavitation induced by an ultrasonic probe. This process has shown much promise in forming both finite layer $MoS_2$ [6] as well as $MoS_2$ nanoparticles [7] in a coherent fashion which could approach mass production. In this study, we utilized a much longer time for the ultrasonic process than in previous studies in an attempt to increase yield and break down the $MoS_2$ to the smallest size possible. However, unlike these previous studies in which pure nanoscale $MoS_2$ was formed, the longer sonication time used here resulted in nanoscale finite layer $MoS_2$ sheets as well as byproducts in the form of multi-walled carbon fullerenes, also known as carbon onions, and finite layer nanoscale graphite platelets. Hence, there appears to be a limit to how finely $MoS_2$ can be broken down using ultrasonic agitation, with longer sonication times leading to sufficient breakdown of the organic solvent to form of carbon based nanoparticle impurities. Unlike the current study, carbon onions are normally only produced using high temperature techniques. This includes irradiation of carbon soot with a high energy electron beam [8-10], annealing of diamond nanoparticles [11], and underwater arc discharge between graphitic electrodes [12]. Interestingly, one study which also used room temperature ultrasonic agitation also found that carbon onions could be formed from multiwalled nanotubes under certain conditions [13]. Our results show that extended sonication is not be optimal for generating pure finite layer MoS2, but could be of interest for generating alternative carbon based nanoparticles such as the carbon onions, which are themselves of interest for applications such as the creation of novel supercapacitors [14].

## 2. Materials and Methods

Sonication was performed with a Vibracell VDX-505 using a 20 kHz solid titanium alloy probe. The reaction vessel was a 400 mL Berzelius beaker cooled using a home-built system consisting of an outer-beaker water jacket fitted with a copper radiator coil. Heat was drawn away from the system by circulating 4°C water through the copper tubing with a Fisher Scientific Isotemp 3006S pump-driven chiller. The $MoS_2$ powder, 300-mesh 99.9% Molybdenum (IV) sulfide from Alfa Aesar, was ball milled prior to use to achieve a fine powder. After ball milling, the powder attained a slight greenish hue indicating the presence of nanoscale $MoS_2$. For ultra-sonication, the processed $MoS_2$ powder was added to the beaker filled with Isopropyl alcohol (99.95%) using a starting concentration of approximately 6.0 g/L. Isopropanol was chosen as it is a readily available commercial solvent, has a low evaporation rate in ambient conditions, and is relatively non-toxic compared to other solvents. The mixture was subjected to ultrasonic cavitation until an arbitrarily chosen energy level had been transferred to the sample. An energy transfer, calculated as a summation of the sonicator output registry (in Joules), of 15 MJ served as the endpoint for cavitation. A pulse mode was chosen for cavitation with a 75% duty cycle operating at 500 W. The duty cycle and power output were chosen after a series of experiments to optimize power transfer and probe life. Actual power transfer was monitored by the system and steadily decreased during the multiple days necessary to achieve 15 MJ due to wear on the ultrasonic probe. Probe wear left micron scale titanium impurities within the sample which were easily removed by centrifuging the sample.

The sonicated mixture was centrifuged in a Fisher Scientific accuSpin Micro 17 at varied times and forces (see Fig.1). The supernatant was removed leaving the larger particulate matter behind. The resulting suspension was optically analyzed for absorbance properties using an Avaspec 2048 USB spectrometer, illuminated with an AvaLight DHc combination deuterium-hydrogen LED source. Spectra were normalized to pure isopropanol. Transmission electron microscopy (TEM) images and energy dispersive x-ray spectroscopy (EDS) were acquired using a JEOL JEM-2100F TEM. A drop cast method was used to prepare samples for TEM using a carbon coated gold grid. A similar drop cast method was used to prepare samples for scanning tunneling microscopy (STM) measurements using a substrate consisting of a 5 nm thick Au film thermally evaporated onto a Si(001) prepared similarly as in previous experiments [15]. After the isopropanol alcohol evaporated, samples were immediately inserted into the chamber of a modified variable temperature UHV STM system (Omicron) with a base pressure of $3\times10^{-9}$ mbar. Images were obtained at room temperature using the constant current mode with tips electrochemically etched from 0.25 mm polycrystalline tungsten wire in a 5 mol KOH solution with a 5 $V_{rms}$ bias.

## 3. Results and Discussion

The effects of ultrasonic agitation were obvious upon visual inspection. The $MoS_2$ suspension transformed from a gray-black color to one with a yellow-green hue. Absorption measurements (Fig. 1) show the differences seen in samples exposed to different centrifuge processes. Five peaks characteristic of $MoS_2$ are observed in the spectra of each sample and are labeled A, B, C, D, and E. The first two peaks are located at approximately 1.87 (A) and 2.04 eV (B) and correspond to direct exciton transitions at the K point [16, 17]. The energy difference between these two peaks is known to be the result of spin-orbit splitting in the valence band [18, 19]. The location of peaks C and D show some variation with increasing centrifuge rates. Most dramatic occurs as the centrifuge rate is increased from 3g to 6g causing peak C to shift from 2.64 to 2.80 eV. The location of peak C has been previously observed at roughly 2.62 eV in bulk $MoS_2$ and found to transition to about 2.84 eV in monolayer $MoS_2$ [20]. Therefore, the blue shift of these peaks is consistent with the relative reduction in the amount of bulk-like $MoS_2$ particles from the solution with increasing centrifuge rates. Peak E is not often discussed in literature as most optical absorption spectra reported do not extend this far into the UV range. However, similar peaks have been found in the spectra of other suspended $MoS_2$ solutions [5, 21].

A number of less apparent peaks appear in the absorption spectra that are not known to be associated with $MoS_2$ in any form. For example, a small peak appears in the spectra at approximately 3.35 eV most prominent in the solutions centrifuged at 3 and 6g for 60 min, but also appear to lesser extents in the other solutions. A series of TEM measurements were carried out on the different solutions to better understand the source of these unknown states. As can be seen in Fig. 2a, the sample that underwent 17g centrifuge for 99 minutes is composed of thin platelets as might be expected for $MoS_2$ which has been broken down into nanoscale form. These platelets have lateral dimensions ranging from tens to hundreds of nanometers, meaning that the resultant products are truly nanoscale in nature. Consistent with the characteristic $MoS_2$ peaks in the absorption spectra, EDS analysis of the platelets (Fig. 2b) indicate the presence of molybdenum. Localized spikes in Mo concentrations correspond to darker regions in the TEM images and may indicate that the $MoS_2$ balled up during the sonication or drop casting process. Interestingly, EDS also shows that some form of carbon blankets the entire structure.

Although difficult to observe on the platelets, isolated nanoparticles can be seen surrounding the larger features. Upon closer examination (Fig. 3a) these nanoparticles are found to contain several sub-particles with geometric facets and an overall elliptical shape. The lack of a structure in the center is consistent with previous TEM images of multi-walled fullerenes [22-24]. The spacing of the individual lattices (Fig. 3b) shows a separation of 0.34 nm, which is consistent with the lattice spacing of carbon fullerenes and not $MoS_2$. These fullerene particles are not present in the pure isopropanol prior to sonication nor are they found in solutions sonicated with less energy input. Therefore, we conclude they must originate from the breakdown of the isopropanol molecules after long term sonication.

STM analysis of samples centrifuged at 6g and 17g show that the platelets are typically decorated with nanoparticles and in most cases coverage is nearly 100%. The topographic image in Fig. 4a shows a partially exposed platelet from the solution centrifuged at 6g. Nanoparticle heights range between 5 and 10 nm and have lateral dimensions on the order of 10 nm, consistent with the carbon onions observed in in Fig. 3a. Although most exposed platelets exhibit the atomic step heights expected for $MoS_2$, some show step heights consistent with graphite such as the cross section profile in Fig. 4b taken along the dashed line in Fig. 4a. The presence of finite layer graphitic sheets and multi-walled fullerenes would certainly account for the presence of carbon in the EDS data. Research has shown that organic cracking can be catalyzed through the use of $MoS_2$ [25], which could explain how long term sonication of $MoS_2$ in an organic solvent results in the formation of the thin graphite platelets and perhaps the formation of carbon onions as well. The sonication process breaks down the bonds of the organic solvent, allowing carbon to form upon the finite layer $MoS_2$ sheets which are simultaneously broken down from the bulk. Previous experiments have demonstrated that carbon onions have a characteristic absorption peak at 5.7 eV due to surface plasmons which would be outside our data range [26]. The absorption peak associated with graphite nanoplatelets is around 4.59 eV, but is likely overwhelmed by the larger signal from the $MoS_2$ [27].

## 4. Conclusions

Liquid exfoliation is a proven method to create finite layer $MoS_2$. Although increased sonication is perhaps beneficial to produce a higher yield of nanoscale $MoS_2$, it can also lead to the formation of carbon byproducts in the form of finite layer graphite nanoparticles and carbon onions. These carbon structures likely form as isopropanol alcohol molecules are broken apart during sonication, and in our study completely cover any $MoS_2$ found in our samples. The presence of conductive carbon would serve to short out the direct gap which makes nanoscale $MoS_2$ attractive in many cases. However, the ability to create carbon onions using ultrasonic agitation would appear to be unique in that it does not require high temperatures and may hold interest in itself for future methods of synthesis.

**Acknowledgments**

The authors would like to acknowledge use of the University of Iowa Central Microscopy Research Facility, a core resource supported by the Vice President for Research & Economic Development, the Holden Comprehensive Cancer Center and the Carver College of Medicine. The authors would also like to thank Dr. Jian Shao for carrying out the TEM measurements. This work was supported by the National Science Foundation Grant Nos. DMR-1410496, DMR-1206530, and CAREER-DMR-1552482.

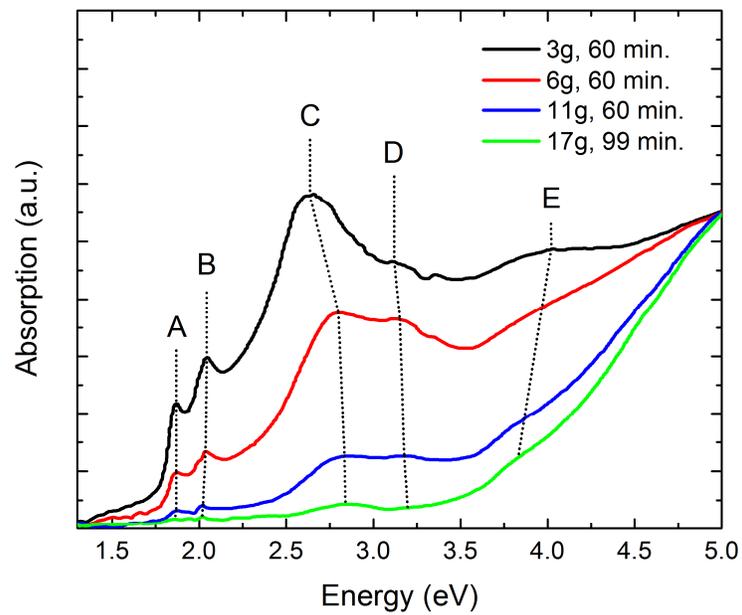

Figure 1. Optical absorption measurements of MoS$_2$ suspensions after 15 MJ of ultrasonic agitation after various centrifuge conditions. Exciton peaks characteristic of MoS$_2$ are labeled on the figure. Resultant spectra were normalized to absorption at 5 eV for comparison.

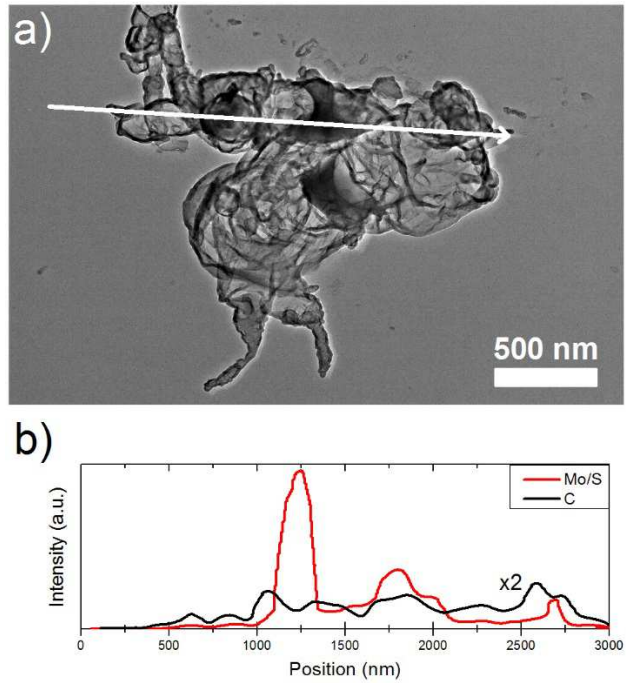

Figure 2. TEM and local EDS analysis of the sample extracted from the supernatant after 17g centrifuge rate for 99 minutes. a) TEM image showing platelets commonly seen throughout the sample. b) EDS analysis showing relative intensity related the C Kα and the overlapping Mo Lα and S Kα peaks along the line shown in panel (a).

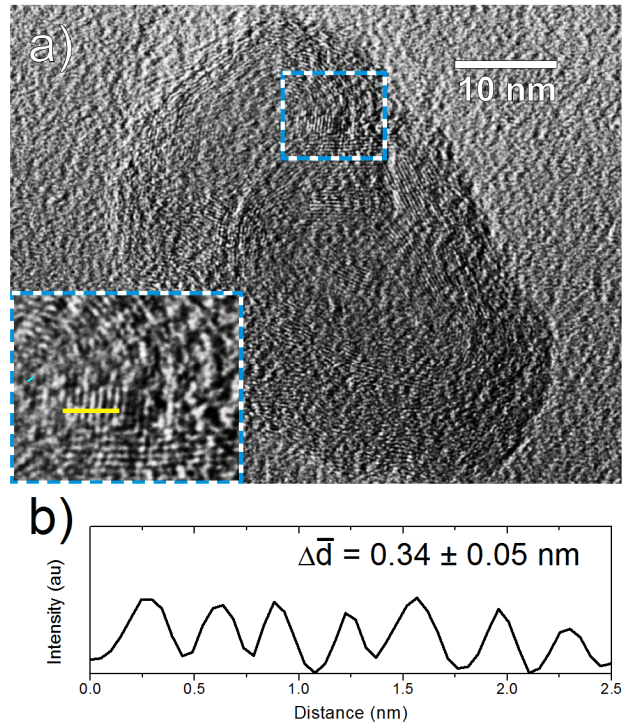

Figure 3. TEM analysis of the smaller nanoparticles in the sample centrifuged at 17g for 99 minutes. a) Nanoparticles with inset showing the atomic lattice. b) Profile along the yellow line in (a) showing an average spacing of 0.34 nm.

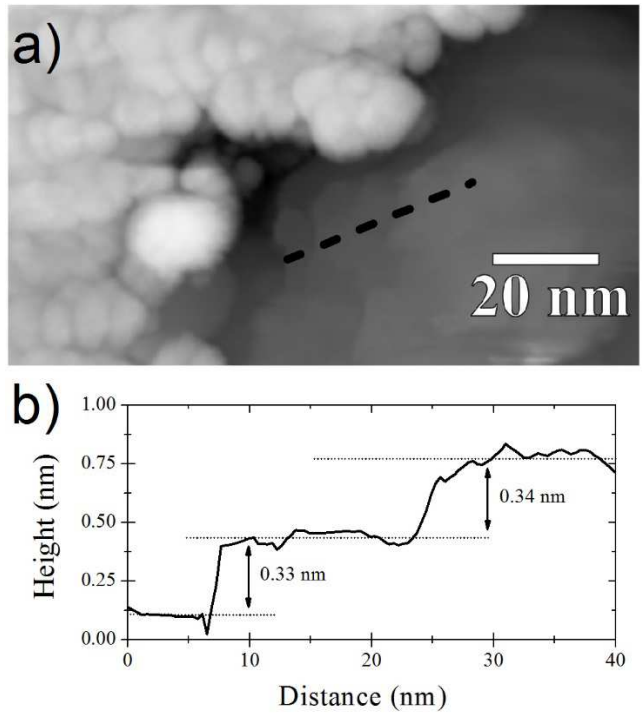

Figure 4. STM data taken from the sample centrifuged at 6g for 99 minutes. a) Image of an exposed graphite platelet partially covered in carbon nanoparticles. b) Cross section profile taken along the dashed line in (a).

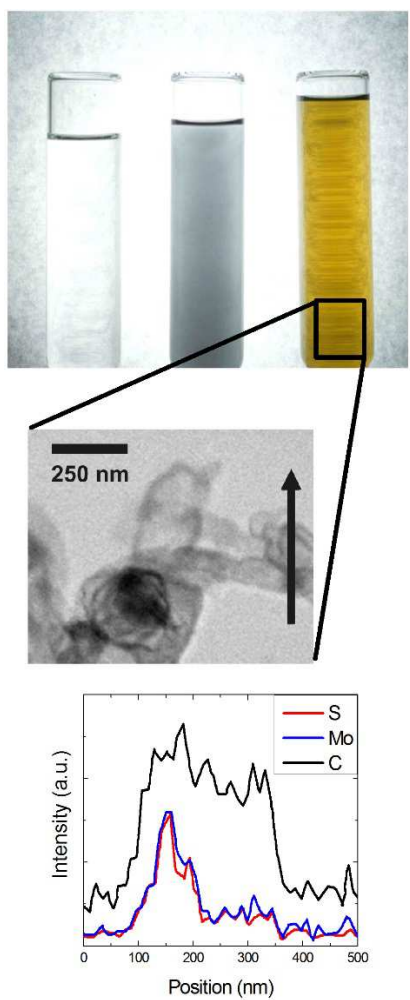

Graphical Abstract: Optical photograph of (from top to bottom) pure isopropanol, ball-milled MoS$_2$ powder suspended in isopropanol, and the extracted supernatant from a sample subjected to 15 MJ of ultrasonic agitation and centrifuged at 3 g for one hour. TEM images show expected formation of finite layer material. However, EDS indicates a substantial amount of carbon incorporated into the finite layer material.